\documentclass[epj]{svjour}

\usepackage[autostyle,italian=guillemets]{csquotes}
\usepackage[utf8]{inputenc}
\usepackage[nowrite,infront,standard]{frontespizio}
\usepackage[english]{babel}
\usepackage{physics}
\usepackage{amsmath,amssymb}
\usepackage [svgnames]{xcolor}
\usepackage{mathrsfs}
\usepackage{subfloat}
\usepackage{frontespizio}
\usepackage{chemmacros}
\usepackage{float}
\usepackage{graphicx} 
\usepackage{bm}
\usepackage{appendix}
\usepackage{hyperref}
\usepackage{midpage}
\usepackage{placeins}
\usepackage{subfig}

\begin{document}
\title{Nonlocality and entropic uncertainty relations in neutrino oscillations}
\author{Massimo Blasone\inst{1,2}, Silvio De Siena\inst{3} and Cristina Matrella\inst{1,2} 
}                     
%
%
\institute{ Dipartimento di Fisica, Universit\`a degli Studi di Salerno, Via Giovanni Paolo II, 132 84084 Fisciano, Italy \and INFN, Sezione di Napoli, Gruppo Collegato di Salerno, Italy \and Retired Professor, Universit\`a degli Studi di Salerno,  email: silvio.desiena@gmail.com}
\date{Received: date / Revised version: date}
%
\abstract{
Using the wave-packet approach to neutrino oscillations, we analyze Quantum-Memory-Assisted Entropic Uncertainty Relations  and show that  uncertainty and  the Non-local Advantage of Quantum Coherence are anti-correlated. Furthermore, we explore the hierarchy among three different definitions of NAQC, those based on the $l_{1}$-norm, relative entropy and skew information coherence measures, and we find that the coherence content detected by the $l_{1}$-norm based NAQC  overcomes  the other two. The connection between QMA-EUR and NAQC could provide a better understanding of the physical meaning of the results so far obtained, and suggest a their extension to quantum field theory.
\PACS{
      {PACS-key}{discribing text of that key}   \and
      {PACS-key}{discribing text of that key}
     } 
} 
\authorrunning{M.~Blasone, S.~De~Siena and C.~Matrella}
\maketitle
\section{Introduction}

Coherence and quantum correlations play a central role in quantum information, quantum computation and cryptography \cite{Streltsov}, and recently the extension to particle physics of these topics has attracted much attention in view of potential applications in the above cited fields. Different definitions and quantifications of coherence and of quantum correlations have been introduced, and it seems that they can find a natural and clear placing inside the complete complementarity relations (CCR) approach \cite{Basso20}-\cite{NoiCCR}.

During the years, many studies have been carried on quantum correlation in neutrino systems \cite{Blas1}-\cite{PoS}, but only recently  the attention has been placed on the study of uncertainty relation in neutrino oscillations \cite{Wang},\cite{Li}.

The uncertainty principle is one of the cardinal point of quantum mechanics. It provides a limit to our ability to accurately predict the measurement outcomes for a pair of incompatible observables of a quantum system. The  principle was expressed for the first time by Heinsenberg, in 1927, for position and momentum \cite{Heinsenberg} and  was formulated by Kennard \cite{Kennard} in terms of variances. Afterwards, it was generalized to any pair of incompatible observables and with the advent of Shannon information theory, the uncertainty relations have been expressed in terms of entropy.
The first to wonder if the uncertainty principle could be expressed in terms of entropy was Everett \cite{Everett}, who found Hirschman's support \cite{Hirschman} and over the years many studies have addressed this topic \cite{Beckner}-\cite{Maassen}.

Recently, advances have been made that allow a generalization of such studies to the case in which the parts of the considered system  can be correlated  in a non-classical way \cite{Berta}-\cite{Coles}. In fact,  the correlation between subsystems  can be used to reduce the  uncertainty below the usual limits,
and the entropic uncertainty relations has been generalized to the case of  quantum memory -- Quantum-Memory-Assisted Entropic Uncertainty Relations (QMA-EUR). 

In this paper, we consider the QMA-EUR in the physical context of neutrino oscillations. Moreover, to  investigate what happens when correlations between  systems are present, we consider  the Non-local Advantage of Quantum Coherence (NAQC) \cite{Mondal}, which has been proved to be the strongest one inside the whole hierarchy of quantum correlations, overtaking even the Bell non-locality. NAQC occurs in a bipartite system when the average coherence of the conditional state of a subsystem B, after a local measurements on A, exceeds the coherence limit of the single subsystem. Several definitions of NAQC have been formulated which differ in the coherence measures adopted. 

In this work, we analyze the relation between NAQC and uncertainty, using the relative entropy as coherence measure. Then, we note that the three criteria $N_{\alpha}(\rho_{AB})>C_{\alpha}$ corresponding to  $\alpha=l_{1},re,sk$ (respectively $l_{1}$-norm, relative entropy and skew information) may identify different regions of NAQC correlation, so NAQC is not only related to the form of the quantum state, but also to the choice of coherence measures. The question is if there exists a hierarchy between these three different NAQC definitions in the case of neutrino oscillations and, assuming the value of the physical parameters of three different experiments, we provide a positive answer.

At last, we want to remark that uncertainty is directly related to observational aspects, and in particular to variances. Therefore, the connection with NAQC, highlighted in Ref.\cite{Wang} and in this paper, could on the one hand  clarify the physical meaning of the results obtained about in neutrino oscillations \cite{NoiCCR},\cite{Us},\cite{PoS}, on the other hand, relying on the approach of \cite{BV1,BV2}, help to extend this results to quantum field theory in terms of variances of hermitian operators (flavor changes), analogously to what done in Ref.\cite{EPL}.
The plan of the paper is as follow: in Section 2 we recall the notions of QMA-EUR and of NAQC and find their expression in terms of neutrino oscillation probabilities. We also investigate the relation between the NAQC and uncertainty using a wave packet approach, finding that the uncertainty is anti-correlated with the NAQC. In Section 3, we further analyze the NAQC in neutrino oscillations. It comes out that in this context a hierarchy between the three different NAQC is present. Section 4 is devoted to conclusions and outlook.

\section{QMA-EUR and NAQC}
\label{S2}
Let us suppose Bob prepares a bipartite system $\rho_{AB}$, where the two parts A and B are correlated. Then he sends part A to Alice and keeps part B as a quantum memory.  After receiving A, Alice operates on it deciding to measure one of the observables P and R, and tells to Bob her choice. Based on  Alice's measurement choice, Bob is able to guess her outcomes with minimal deviation limited by the uncertainty's lower bound by means of the part B which is correlated with A.

The quantum memory can reduce the uncertainty, and the usual EUR, $H(P)+H(R)\ge-\log_{2}c$, expressed in terms of Shannon entropy \cite{Coles}, is modified, in terms of von-Neumann entropy, as:
\begin{equation}
S(P|B)+S(R|B)\ge-\log_{2} c(P|R)+S(A|B).
\label{1}
\end{equation}
The quantities in Eq.(\ref{1}) are the following:

\begin{itemize}
\item  $S(A|B)=S(\rho_{AB})-S(\rho_{B})$ is the conditional von Neumann entropy of systemic state $\rho_{AB}$ with $S(\rho_{AB})=-tr(\rho_{AB}\log_{2}\rho_{AB})$,
\item  $\rho_{B}=tr_{A}(\rho_{AB})$,
\item   $S(X|B)=S(\rho_{XB})-S(\rho_{B})$ is the conditional von Neumann entropy of $\rho_{XB}=\sum_{i}(|\psi_{i}^{X}\rangle_{A}\langle\psi_{i}^{X}|\mathcal{I}_{B})\rho_{AB}(|\psi_{i}^{X}\rangle_{A}\langle\psi_{i}^{X}|\mathcal{I}_{B})$ (that is the state of B after performing a measurement on A of the observable X with eigenstates $|\psi_{i}^{X}\rangle$), 
\item  $c(P|R)=\max_{j,k}|\langle\psi_{j}^{P}|\phi_{k}^{R}\rangle|^{2}$ represents the maximal overlap between the eigenstates $|\psi_{j}^{P}\rangle$ and $|\phi_{k}^{R}\rangle$ of the observables P and R.
\end{itemize}

 Since the correlations reduce uncertainty and the NAQC represents the strongest  quantifier of quantum correlations, we verify its role within QMA-EUR. 
 Several definitions of NAQC have been provided, based on different coherence measures.
Given a state $\rho$ in the reference basis $\{\ket{i}\}$, a measure of coherence takes the form:

\begin{equation}
C_{D}(\rho)=\min_{\delta\in I}D(\rho,\delta),
\label{2}
\end{equation}
that is the minimum distance between $\rho$ and the set of incoherent states $I$. $D(\rho,\delta)$ is a distance measure between two quantum states. For example, one can consider $D(\rho,\delta)=||\rho-\delta||_{l_{1}}$, with $||.||_{l_{1}}$ is the $l_{1}$-norm or $D(\rho,\delta)=S(\rho||\delta)$, the quantum relative entropy. By minimizing over the set of incoherent states, one can obtain two bona fide measures of coherence \cite{Braumgratz} as:

\begin{equation}
C_{l_{1}}(\rho)=\sum_{i\ne j}|\bra{i}\rho\ket{j}|, \hspace{1cm} C_{re}(\rho)=S(\rho_{diag})-S(\rho),
\label{3}
\end{equation}
where $S(\rho)$ is the von Neumann entropy of $\rho$ and $\rho_{diag}$ is the matrix of the diagonal elements of $\rho$.\\

Another possible coherence measure we can consider is the skew information \cite{Girolami}:

\begin{equation}
C_{sk}(\rho)=-\frac{1}{2}Tr\{[\sqrt{\rho},K]^{2}\} ,
\label{3.1}
\end{equation}
where $K$ is an operator.

Mondal et al. \cite{Mondal} defined the NAQC of a bipartite state $\rho_{AB}$ considering the average coherence of the post measurement state $\{ p_{B|\Pi_{i}^{a}},\rho_{B|\Pi_{i}^{a}}\}$ of B after a local measurement $\Pi_{i}^{a}$ on A:

\begin{equation}
N_{\alpha}(\rho_{AB})=\frac{1}{2}\sum_{i\ne j, a =\pm} p_{B|\Pi_{i}^{a}}C_{\alpha}^{\sigma_{j}}(\rho_{B|\Pi_{i}^{a}}),
\label{3.2}
\end{equation}
where $\Pi_{i}^{\pm}=\frac{I\pm\sigma_{i}}{2}$, with $I$ and $\sigma_{i},(i=1,2,3)$ being the identity and the three Pauli operators; $ p_{B|\Pi_{i}^{a}}=\Tr(\Pi_{i}^{a} \rho_{AB})$, $\rho_{B|\Pi_{i}^{a}}=\Tr_{A}(\Pi_{i}^{a} \rho_{AB})/ p_{B|\Pi_{i}^{a}}$. $C_{\alpha}^{\sigma_{j}}(\rho_{B|\Pi_{i}^{a}})$ is the coherence of the conditional state of B with respect to the eigenbasis of $\sigma_{j}$, with $\alpha=l_{1},re,sk$. For a one-qubit state $\rho$, the sum of $C_{\alpha}(\rho)$ with respect the three mutually unbiased bases are upper bounded, respectively, by $ C_{l_{1}}=\sqrt{6}$, $C_{re}=2.23$ and $C_{sk}=2$. If $N_{\alpha}(\rho_{AB})> C_{\alpha}$, $\rho_{AB}$ is said to have acquired NAQC. 

\subsection{QMA-EUR and NAQC in neutrino oscillations}
At time $t$ the state for a  two-flavor neutrino of initial flavor $\alpha$ is:
\begin{equation}
\rho_{AB}^{\alpha}(t)=|a_{\alpha\alpha}(t)|^{2}|10\rangle\langle10|+a_{\alpha\beta}(t)a_{\alpha\alpha}^{*}(t)|01\rangle\langle10|
+|a_{\alpha\beta}(t)|^{2}|01\rangle\langle01|+a_{\alpha\alpha}(t)a_{\alpha\beta}^{*}(t)|10\rangle\langle01|
\label{4}
\end{equation}
where $\alpha,\beta=e,\mu,\tau$.
We choose as incompatible observables $(P,R)=(\sigma_{x},\sigma_{y})$. In this case the maximal overlap is $c(P,R)=\frac{1}{2}$.
Starting by Eq.(\ref{4}) (see Appendix \ref{AppendixA}), we find that entropic uncertainty $U^{\alpha}$ and the uncertainty's lower bound $U_{b}^{\alpha}$ are the LHS and the RHS of Eq.(\ref{1}), respectively. They can be expressed in terms of oscillation probabilities \footnote{The expression of neutrino oscillation probability in the wave packet approach is obtained in Appendix \ref{AppendixB}} as:

\begin{equation}
U^{\alpha}(t)=S(\rho_{PB}^{\alpha}(t))+S(\rho_{RB}^{\alpha}(t))-2S(\rho_{B}^{\alpha}(t))
=2(P_{\alpha\alpha}(t)\log_{2}P_{\alpha\alpha}(t)+P_{\alpha\beta}(t)\log_{2}P_{\alpha\beta}(t)+1)
\label{5}
\end{equation}
\begin{equation}
U^{\alpha}_{b}(t)=S(\rho_{AB}^{\alpha}(t))-S(\rho_{B}^{\alpha}(t))-\log_{2}c(P,R)
=P_{\alpha\alpha}(t)\log_{2}P_{\alpha\alpha}(t)+P_{\alpha\alpha}(t)\log_{2}P_{\alpha\beta}(t)+1
\label{6}
\end{equation}

Since the QMA-EUR is expressed in terms of von Neumann entropy, we consider the entropy-based NAQC to investigate the relation between these quantities. The expression of the entropy-based NAQC in terms of oscillation probabilities is:
\begin{equation}
N(\rho_{AB}^{\alpha})(t)=2-P_{\alpha\alpha}(t)\log_{2}P_{\alpha\alpha}(t)-P_{\alpha\beta}(t)\log_{2}P_{\alpha\beta}(t)
\label{7}
\end{equation}

By these equations is simple to obtain the result:
\begin{equation}
U^{\alpha}(t)=2U_{b}^{\alpha}(t)=2[3-N(\rho_{AB}^{\alpha}(t))].
\label{8}
\end{equation}
Eq.(\ref{8}) shows how a stronger quantum correlation will lead to a reduced uncertainty.

\begin{figure}[t]
\centering
 \subfloat[][\emph{DAYA BAY experiment}]{{\includegraphics[width =6 cm]{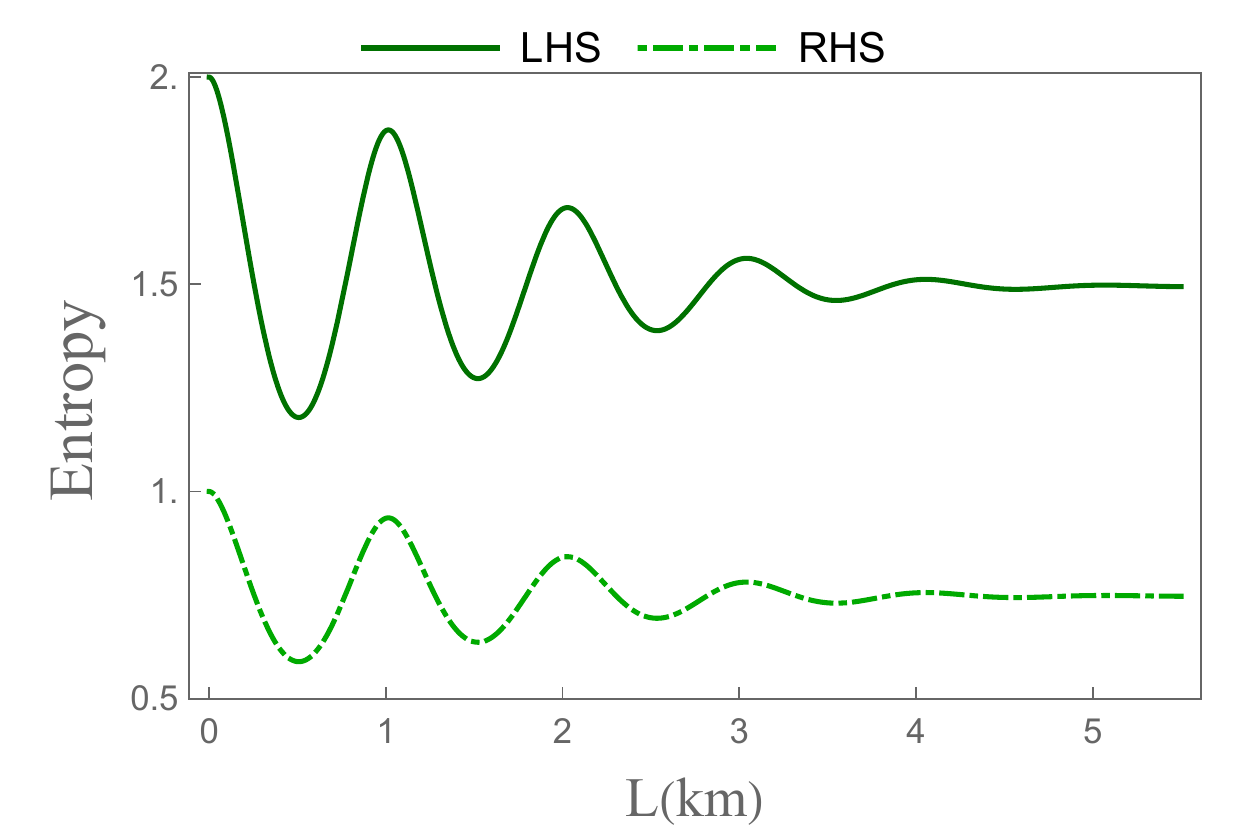}}\quad\quad
{\includegraphics[width =5.7 cm]{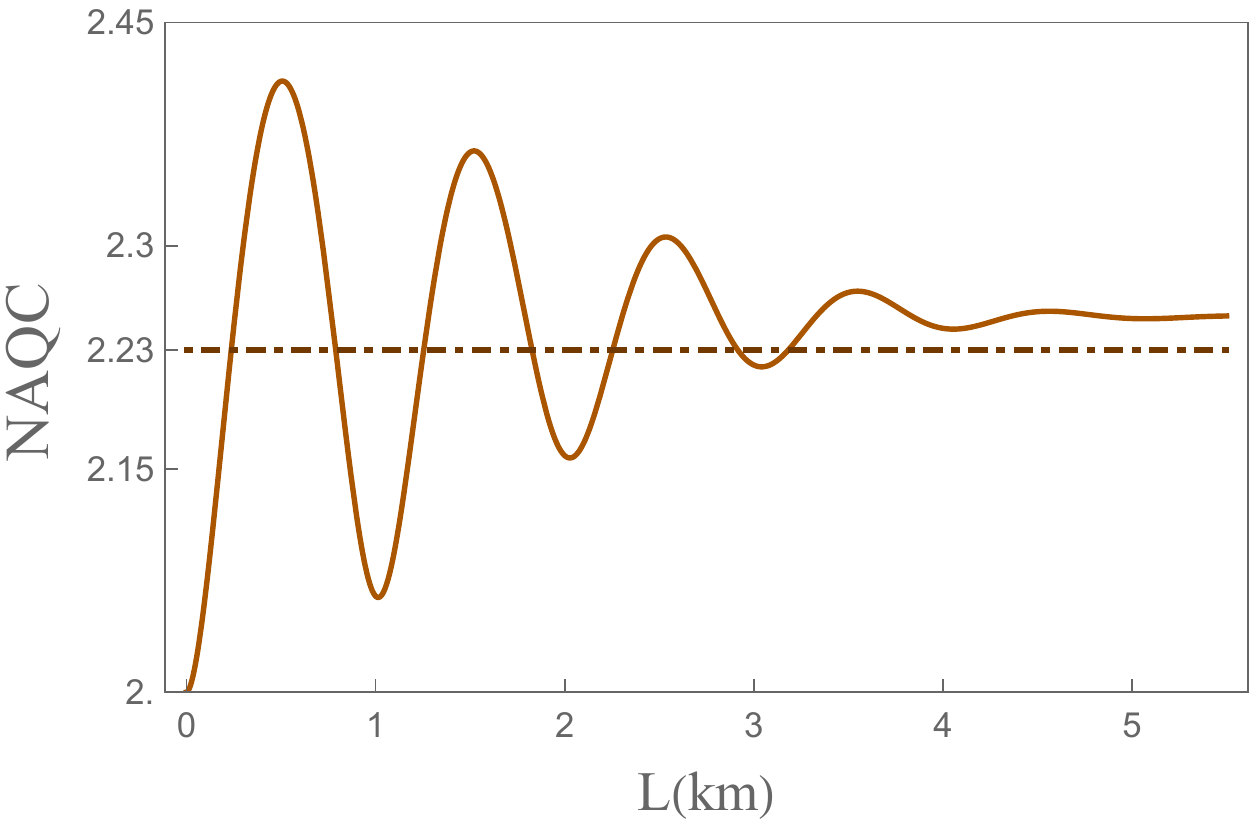}}}\\
 \subfloat[][\emph{KamLAND experiment}]{{\includegraphics[width =6 cm]{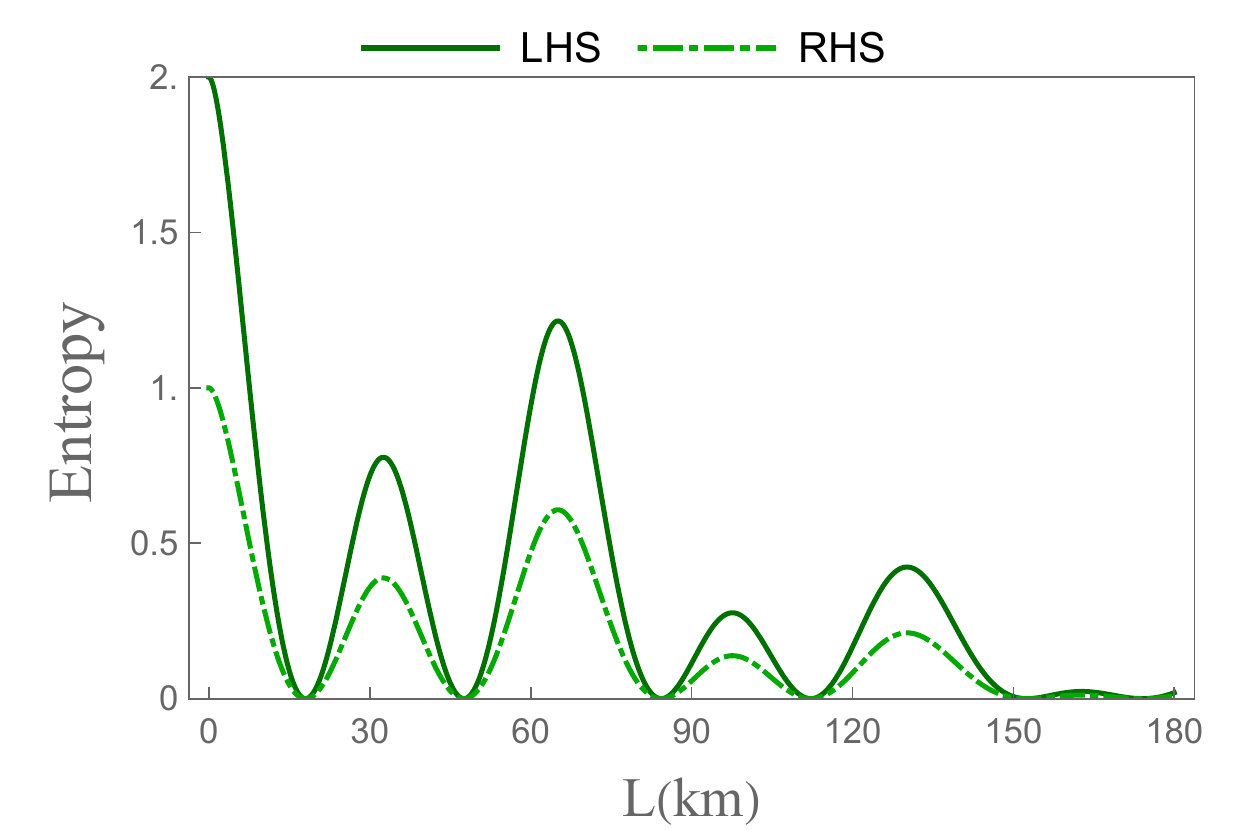}}\quad\quad
{\includegraphics[width =5.7 cm]{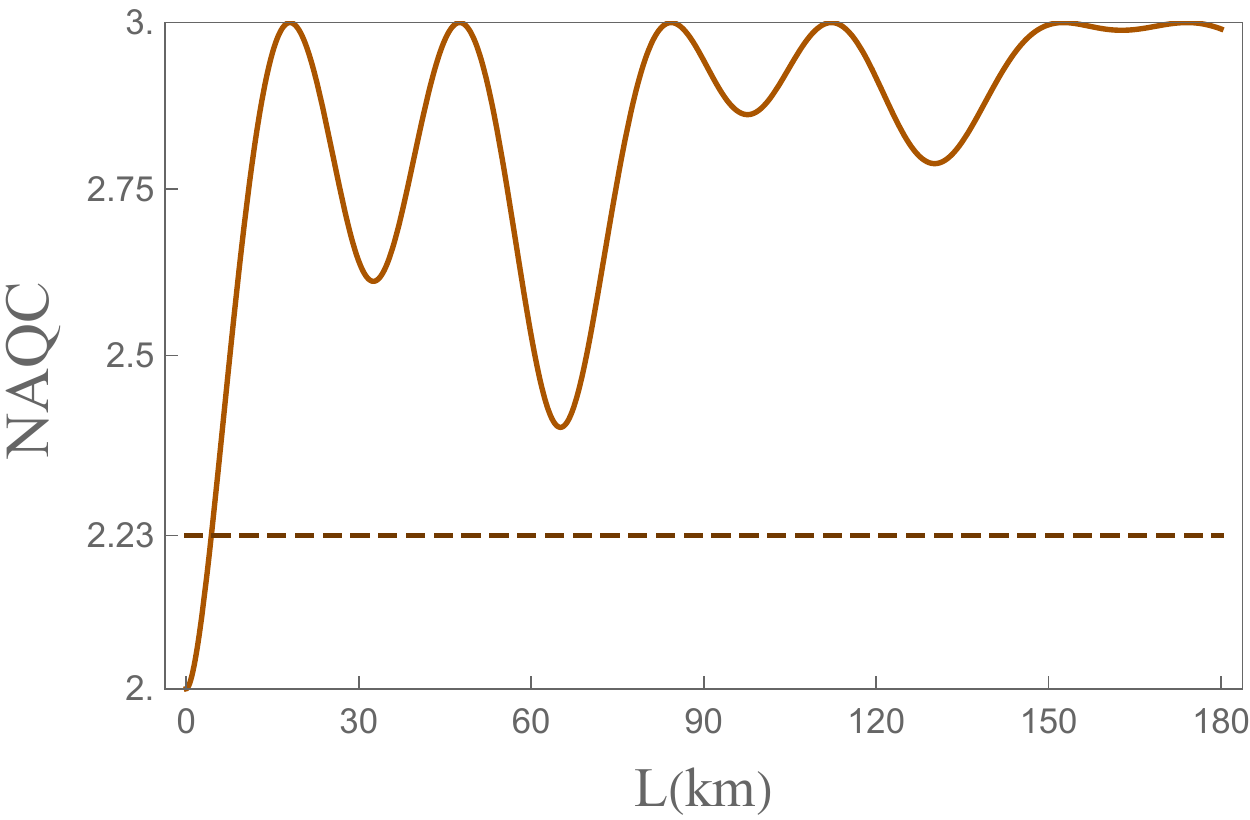}}}\\
 \subfloat[][\emph{MINOS experiment}]{{\includegraphics[width =6 cm]{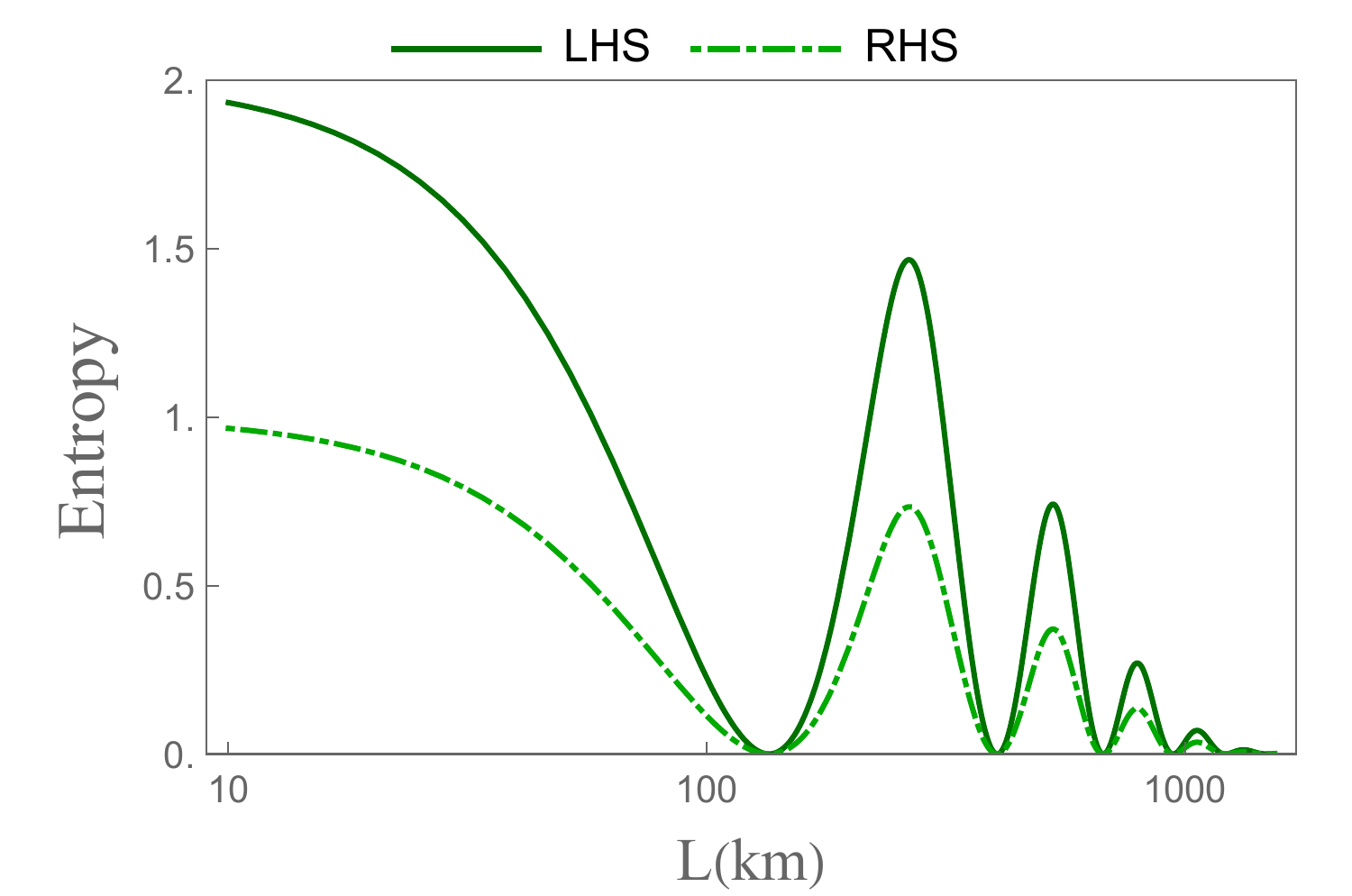}}\quad\quad
{\includegraphics[width =5.7 cm]{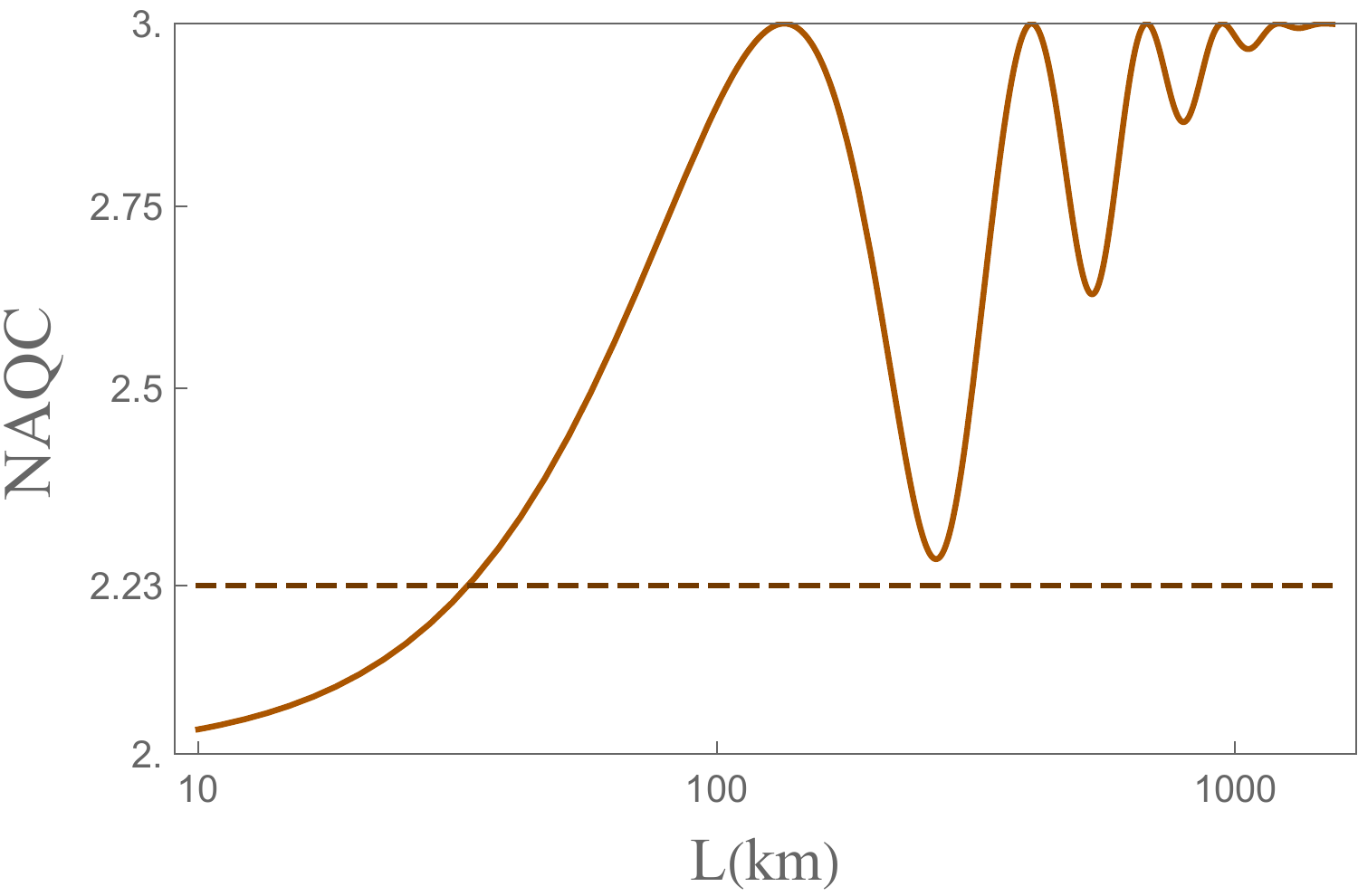}}}
\caption{ QMA-EUR (on the left panel) and NAQC (on the right panel) as a function of the distance for:
(a) Daya Bay,
(b) KamLAND, (c) MINOS.  }
\label{fig1}
\end{figure}

In Fig.(\ref{fig1}), the entropic uncertainty, the uncertainty's lower bound and the entropy-based NAQC are plotted using the parameters from the Daya Bay \cite{Daya},\cite{Daya1}, KamLAND \cite{Kam},\cite{Kam1} and MINOS \cite{Minos},\cite{Minos1} experiments. Daya-Bay and KamLAND are electron-antineutrino disappearance experiments and their oscillations parameters are  $\sin^{2} 2\theta_{13} = 0.084^{+0.005}_{-0.005}$ and $\Delta m_{ee}^{2}=2.42^{+0.10}_{-0.11}\times 10^{-3} eV^{2}$ for Daya-Bay and  $\tan^{2} 2 \theta_{12}=0.47$ and $\Delta m_{12}^{2}=7.49\times 10^{-5}$ for KamLAND. MINOS is a muonic neutrino disappearance experiment and its oscillation parameters are $\sin^{2} 2\theta_{23} = 0.95^{+0.035}_{-0.036}$ and  $\Delta m_{32}^{2}=2.32^{+0.12}_{-0.08}\times 10^{-3} eV^{2}$.

The wave packet approach confirms the results  by Wang et al. \cite{Wang}, obtained  by means a plane-wave approximation,  according to which the uncertainty is completely anti-correlated with the quantum correlations: the stronger the quantum correlation, the smaller the uncertainty. In particular, we have shown in \cite{Us} that in  the wave packet approach the asymptotic trend of the NAQC depends on the mixing angle, with NAQC attaining its maximum value at great distances if the value of the mixing angle overcomes a threshold \footnote{Note that threshold values are different if we consider NAQC based on different coherence measures.}. This happens in KamLAND and MINOS experiments, and consequently we see from Fig.(\ref{fig1}) that the entropy uncertainty and its lower bound go to zero.

\section{Hierarchy among $\alpha$-based NAQCs in neutrino oscillations}
\label{S3}
In this section, we further analyze the NAQC in neutrino oscillations showing some interesting results which arise from the comparison among the NAQCs based on the three different coherence measures mentioned in Section \ref{S2}.

The expression of the entropy based NAQC in terms of oscillation probabilities is given by Eq. (\ref{7}) and analogously, following Ref.\cite {Mondal}, we determine the expressions of the $l_{1}$-norm based NAQC and of the skew information based NAQC:

\begin{equation}
N^{l_{1}}(\rho_{AB}^{\alpha}(t))=2+2\sqrt{P_{\alpha\alpha}(t)P_{\alpha\beta}(t)},
\label{c2}
\end{equation}
and
\begin{equation}
N^{sk}(\rho_{AB}^{\alpha}(t))=2+4P_{\alpha\alpha}(t)P_{\alpha\beta}(t).
\label{c3}
\end{equation}

\begin{figure}[t]
\centering
 \subfloat[][\emph{DAYA BAY experiment}]{\includegraphics[width =5.65 cm]{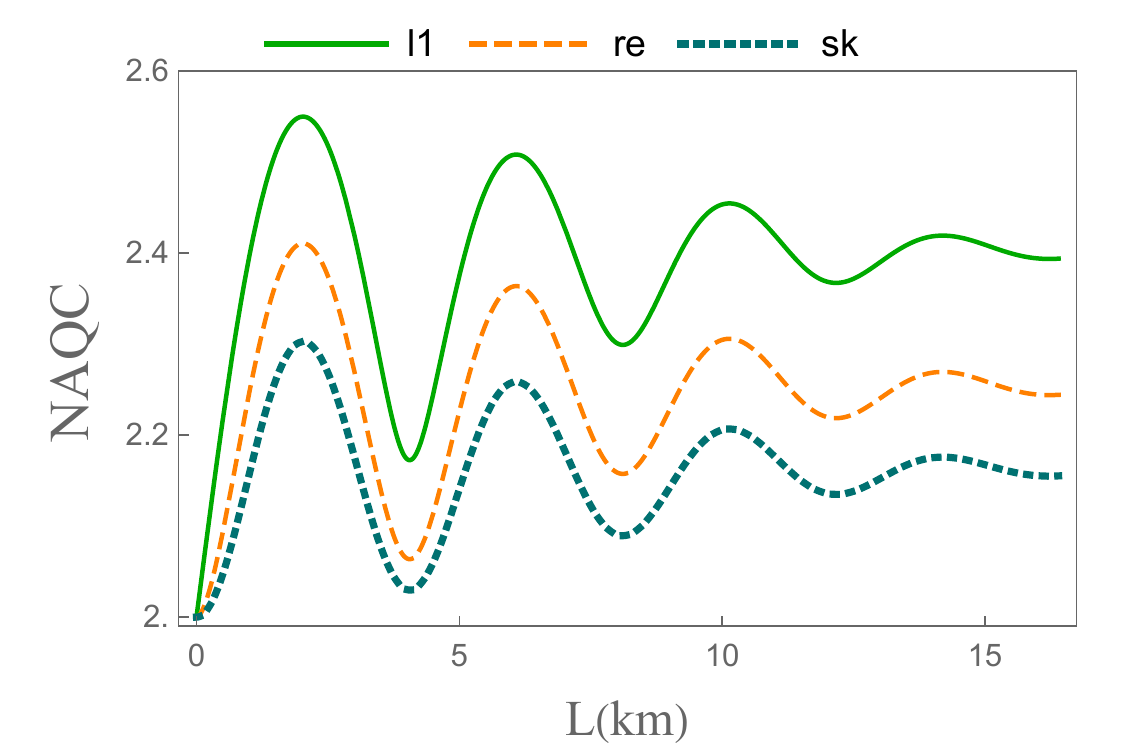}
\label{fig2a}}\quad
 \subfloat[][\emph{KamLAND experiment}]{\includegraphics[width =5.65 cm]{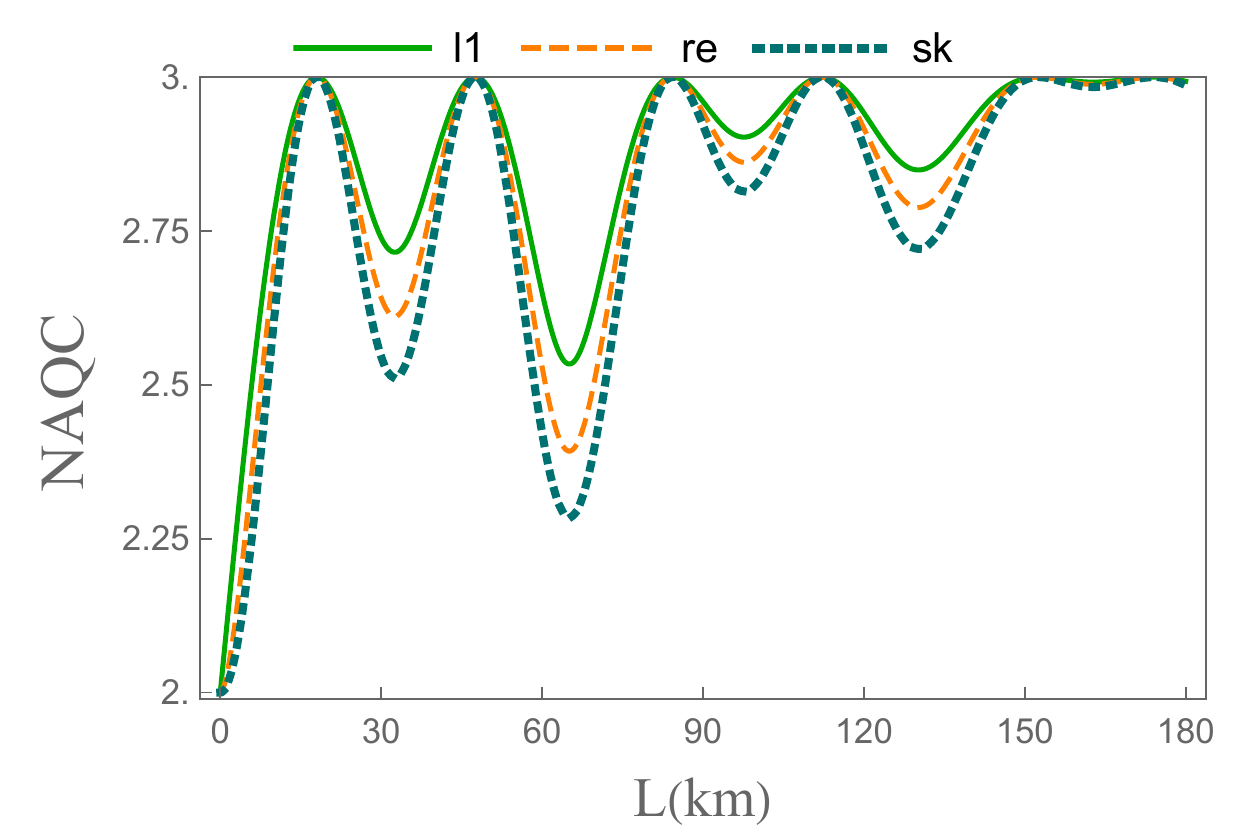}
\label{fig2b}}\quad
 \subfloat[][\emph{MINOS experiment}]{\includegraphics[width =5.65 cm]{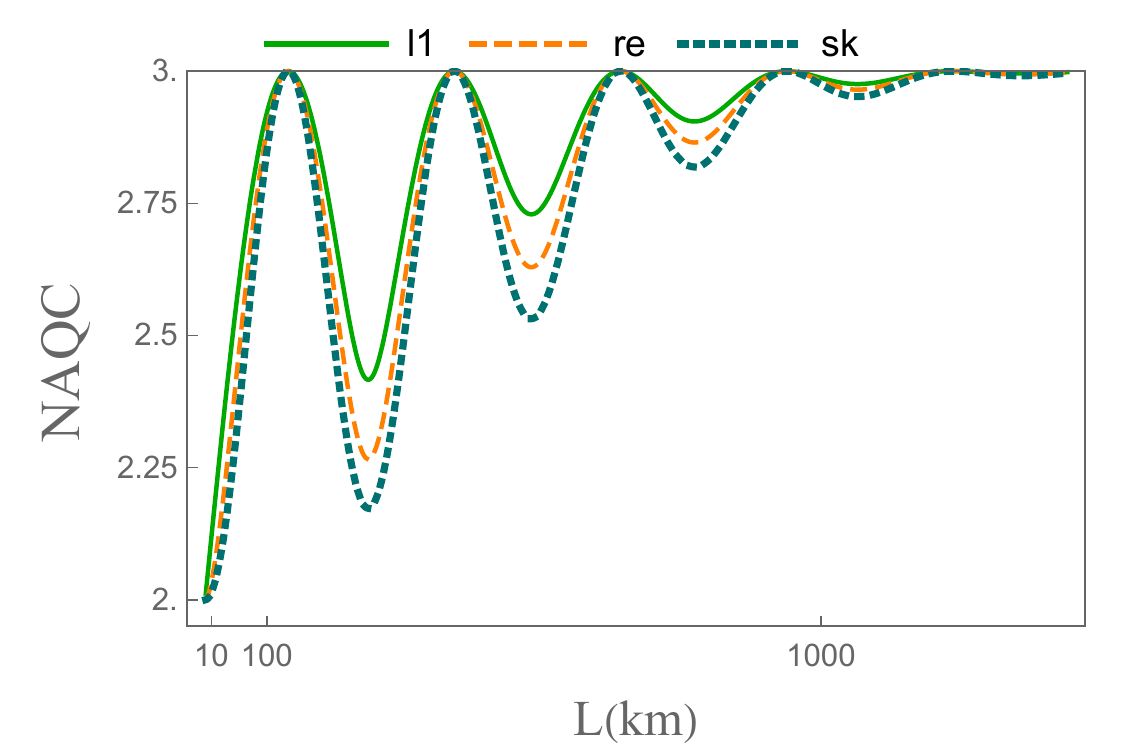}
\label{fig2c}}
\caption{Comparison among the $\alpha$-based NAQCs, $(\alpha=l_{1},re,sk)$ for Daya Bay, KamLAND and MINOS experiments.}
\label{fig2}
\end{figure}

In Fig.(\ref{fig2})  we compare the plots of the NAQC obtained using the three different coherence measures and referring to MINOS, Daya Bay and KamLAND experiments.

\begin{figure}[t]
\centering
\includegraphics[width =8 cm]{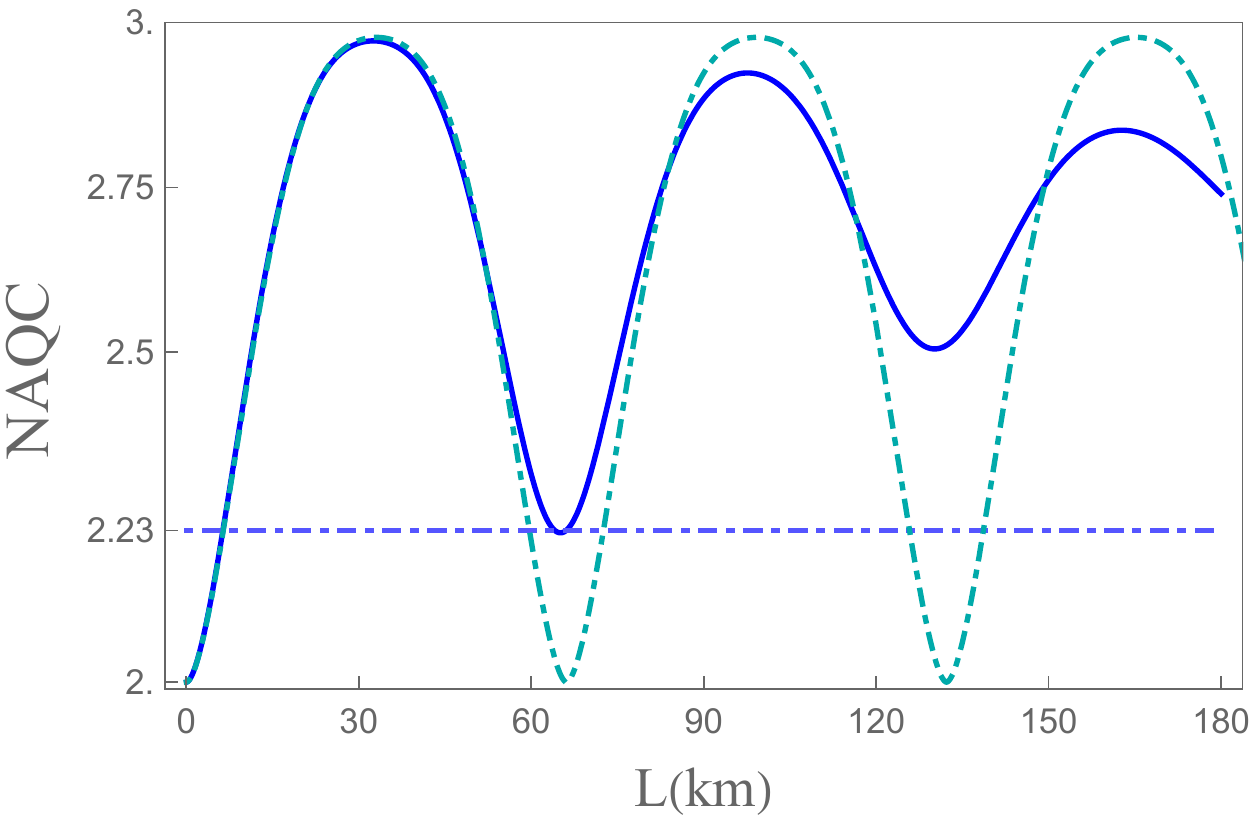}
\caption{Entropy based-NAQC as function of the distance for a mixing angle of $20$ degrees. The solid and the dashed lines represent the plot for the wave packet and plane wave approaches, respectively. }
\label{fig3}
\end{figure}

From Fig.(\ref{fig2}) we can observe how in the case of neutrino oscillations the $l_{1}$-norm  based NAQC is able to capture more quantum resource with respect the others. Furthermore it represents an upper limit for the other two measures.
These observations confirm our previous results about the large distances behavior of NAQC \cite{Us} for all three cases.
In Fig.(\ref{fig2}), we also observe a different trend of the plots relating to the KamLAND experiment compared to the other two. In fact, it can be seen from Fig.(\ref{fig2b}) that there is a non-monotonous growth of the minima of the NAQC. 
In particular, notice how the second minimum is lower than the first. Indeed, in \cite{PoS} we found, in addition to the thresholds determining different asymptotic trends of the NAQCs, a further effect of varying the value of this angle. In an intermediate range of values of this angle, in fact, the monotonicity of NAQC is lost, and this is just the case of KamLAND experiment.
Furthermore, we evaluated the mixing angle for which the minimum of the entropy based-NAQC coincides with the bound value $2.23$, see Fig.(\ref{fig3}). This minimum is positioned at L equal to $\pi$ times the oscillation length and the value of the mixing angle corresponding to it turns out to be $20$ degrees.

 Fig.(\ref{fig3}) also shows a difference between the plots obtained with the plane wave and the wave packet approaches. In fact, by using plane waves we can reach a NAQC for certain regular range of distances. 
This does not happen in the case of wave packets, where above a certain distance we can always reach a NAQC.

\section{Conclusions}
The convergence between topics of quantum information and the physics of elementary particles, in particular of neutrino oscillations, has led in this last field to the development of a wide and systematic study of different definitions and quantifiers of coherence and quantum correlations. In this paper, using a wave-packet approach \footnote{The relevance of the wave packet approach has been discussed in connection with recent experiments in Refs.\cite{DAYAW,JUNO:2021ydg}}, we considered the quantum-memory-assisted entropic uncertainty relation in the physical context of neutrino oscillations, investigating the relation between  the non-local advantage of quantum coherence and uncertainty. We found that the uncertainty is anti-correlated with the NAQC, a result in accordance with what obtained by  Wang et al.  \cite{Wang}, in the  plane-wave approximation.  However, the richer structure generated by the wave-packet approach suggested that entropic uncertainty can go to zero asymptotically in the distance for sufficiently high values of the mixing angle.

Given the several definitions of NAQC based on three different coherence measures – $l_{1}$-norm, relative entropy and skew information – we then proceeded  to further analyze the hierarchy among them. In particular, the $l_{1}$-norm based NAQC resulted to be more able to capture quantum resources with respect the other two quantities. It is obviously interesting to investigate if this hierarchy is confined to the instance of neutrino oscillations, or if it is true in any context. This question will be one subject of future investigations.

Finally, we reiterate what was observed in the introduction, i.e. that the anti-correlation between QMA-EUR and NAQC  could represent a  useful basis for  a better understanding of the physical meaning of the various results recently obtained on NAQC in neutrino oscillations \cite{NoiCCR,PoS,Us}. Indeed, uncertainty relations could be generally formulated in terms of variances and thus are potentially more connected to quantities with an operational meaning, also in view of possible implementation of quantum protocols with neutrinos. We also envisage that the extension of the present results to quantum field theory, which we plan to perform relying on the approach of \cite{BV1,BV2}, could be carried out in terms of variances of  hermitian operators (flavor charges), analogously to what done in Ref.\cite{EPL}.

\newpage
\begin{appendices}
\section{QMA-EUR and NAQC in terms of neutrino oscillation probability}
\label{AppendixA}

We consider the state for a neutrino of flavor $\alpha$:
\begin{equation}
\ket{\nu_{\alpha}(t)}=a_{\alpha\alpha}(t)\ket{\nu_{\alpha}}+a_{\alpha\beta}(t)\ket{\nu_{\beta}}.
\label{a1}
\end{equation}
 
The corresponding density matrix in the orthonormal basis $\{ \ket{00},\ket{01},\ket{10},\ket{11}\}$ is given by Eq.(\ref{4}):

\begin{equation}
\rho^{\alpha}_{AB}(t)=
\begin{pmatrix}
0&0&0&0\\
0&|a_{\alpha\beta}(t)|^{2}&a_{\alpha\beta}(t)a_{\alpha\alpha}^{*}(t)&0\\
0&a_{\alpha\alpha}(t)a_{\alpha\beta}^{*}(t)&|a_{\alpha\alpha}(t)|^{2}&0\\
0&0&0&0
\end{pmatrix}
\label{a2}
\end{equation}

\subsection{QMA-EUR}

We  rewrite Eq.(\ref{1}) for this density matrix, by considering as incompatible observables P and R  the Pauli matrix $\sigma_{x}$ and $\sigma_{y}$. So we have to evaluate:

\begin{equation}
S(\sigma_{x}|B)+S(\sigma_{y}|B)\ge-\log_{2} c(\sigma_{x}|\sigma_{y})+S(A|B)
\label{a3}
\end{equation}
 where:

\begin{equation}
\begin{split}
S(\sigma_{x}|B)&=S(\rho_{\sigma_{x} B})-S(\rho_{B})\\
S(\sigma_{y}|B)&=S(\rho_{\sigma_{y} B})-S(\rho_{B})
\end{split}
\label{a4}
\end{equation}
with:

\begin{equation}
\rho_{\sigma_{x} B}=\sum_{i=1,2}(\ket{x_{i}}\bra{x_{i}})\rho_{AB}(\ket{x_{i}}\bra{x_{i}})=\frac{1}{2}
\begin{pmatrix}
|a_{\alpha\alpha}(t)|^{2}&0&0&a_{\alpha\beta}(t)a_{\alpha\alpha}^{*}(t)\\
0&|a_{\alpha\beta}(t)|^{2}&a_{\alpha\alpha}(t)a_{\alpha\beta}^{*}&0\\
0&a_{\alpha\beta}(t)a_{\alpha\alpha}^{*}(t)&|a_{\alpha\alpha}(t)|^{2}&0\\
a_{\alpha\alpha}(t)a_{\alpha\beta}^{*}&0&0&|a_{\alpha\beta}(t)|^{2}
\end{pmatrix}
\label{a5}
\end{equation}

\begin{equation}
\rho_{\sigma_{y} B}=\sum_{i=1,2}(\ket{y_{i}}\bra{y_{i}})\rho_{AB}(\ket{y_{i}}\bra{y_{i}})=\frac{1}{2}
\begin{pmatrix}
-|a_{\alpha\alpha}(t)|^{2}&0&0&-a_{\alpha\beta}(t)a_{\alpha\alpha}^{*}(t)\\
0&|a_{\alpha\beta}(t)|^{2}&-a_{\alpha\alpha}(t)a_{\alpha\beta}^{*}&0\\
0&-a_{\alpha\beta}(t)a_{\alpha\alpha}^{*}(t)&|a_{\alpha\alpha}(t)|^{2}&0\\
-a_{\alpha\alpha}(t)a_{\alpha\beta}^{*}&0&0&-|a_{\alpha\beta}(t)|^{2}
\end{pmatrix}
\label{a6}
\end{equation}

$\ket{x_{i}}$ and $\ket{y_{i}}$ are the eigenstates of $\sigma_{x}$ and $\sigma_{y}$, respectively. 
$\rho_{B}$ is given by the partial trace with respect A of $\rho_{AB}$:

\begin{equation}
\rho_{B}=
\begin{pmatrix}
|a_{\alpha\alpha}(t)|^{2}&0&\\
0&|a_{\alpha\beta}(t)|^{2}
\end{pmatrix}
\label{a7}
\end{equation}

By considering the spectral representation $\rho=\sum_{i}\lambda_{i}\ket{i}\bra{i}$, we can evaluate the von Neumann entropy as $S(\rho)=-\sum_{i}\lambda_{i}\log_{2}\lambda_{i}$, where $\lambda_{i}$ are the eigenvalues of $\rho$. Hence, we find $S(\rho_{\sigma_{x} B})=-2\cdot\frac{1}{2}(|a_{\alpha\alpha}(t)|^{2}+|a_{\alpha\beta}(t)|^{2})\log_{2}[\frac{1}{2}(|a_{\alpha\alpha}(t)|^{2}+|a_{\alpha\beta}(t)|^{2})]$. However $|a_{\alpha\alpha}(t)|^{2}=P_{\alpha\alpha}$ and $|a_{\alpha\beta}(t)|^{2}=P_{\alpha\beta}$ are nothing more than survival and transition probability, respectively, such that $P_{\alpha\alpha}+P_{\alpha\beta}=1$. Thus  $S(\rho_{\sigma_{x} B})=1$. We also find $S(\rho_{B})=-P_{\alpha\alpha}\log_{2}P_{\alpha\alpha}-P_{\alpha\beta}\log_{2}P_{\alpha\beta}$. From Eq.(\ref{a4}) it is clear that $S(\sigma_{x}|B)=1+P_{\alpha\alpha}\log_{2}P_{\alpha\alpha}+P_{\alpha\beta}\log_{2}P_{\alpha\beta}$. In the same way we can evaluate $S(\sigma_{y}|B)=S(\sigma_{x}|B)$, $S(A|B)=P_{\alpha\alpha}\log_{2}P_{\alpha\alpha}+P_{\alpha\beta}\log_{2}P_{\alpha\beta}$ and $c(\sigma_{x},\sigma_{y})=\frac{1}{2}$.

At this point we are able to write the left hand side and the right hand side of Eq.(\ref{a3}), corresponding respectively to the entropic uncertainty and to the uncertainty's lower bound as:

\begin{equation}
U^{\alpha}=2(P_{\alpha\alpha}\log_{2}P_{\alpha\alpha}+P_{\alpha\beta}\log_{2}P_{\alpha\beta}+1)
\label{a8}
\end{equation}

\begin{equation}
U^{\alpha}_{b}=P_{\alpha\alpha}\log_{2}P_{\alpha\alpha}+P_{\alpha\alpha}\log_{2}P_{\alpha\beta}+1
\label{a9}
\end{equation}

\subsection{NAQC}
Here we show  the basic steps to write the NAQC in terms of neutrino probability. 

Following \cite{Mondal} we decompose our density matrix, Eq.(\ref{4}), as:
\begin{equation}
\rho_{AB}=\frac{1}{4}\bigl(I_{4}+\vec{r}\cdot\vec{\sigma}\otimes I_{2}+I_{2}\otimes \vec{s}\cdot\vec{\sigma}+\sum_{i,j}t_{ij}\sigma_{i}\otimes\sigma_{j}\bigl),
\label{n1}
\end{equation}
where $\vec{r}\equiv (r_{x},r_{y},r_{z})$, $\vec{s}\equiv (s_{x},s_{y},s_{z})$ and $t_{ij}$ are the correlation matrix elements. The decomposition coefficients can be found as: $r_{i}=\Tr[\rho_{AB}(\sigma_{i}\otimes I_{2})]$, $s_{i}=\Tr[\rho_{AB}(I_{2}\otimes\sigma_{i})]$ and $t_{ij}=\Tr[\rho_{AB}(\sigma_{i}\otimes\sigma_{j})]$ , $(i,j=x,y,z)$, where $\sigma_{i}$ are the Pauli matrices. We need these coefficients to evaluate Eq.(\ref{3.2}), i.e:

\begin{equation}
N_{\alpha}(\rho_{AB})=\frac{1}{2}\sum_{i\ne j, a =\pm} p_{B|\Pi_{i}^{a}}C_{\alpha}^{\sigma_{j}}(\rho_{B|\Pi_{i}^{a}}),
\label{n2}
\end{equation}
Here, $p_{B|\Pi_{i}^{a}}=\frac{1}{2}\gamma_{ja}$ and  $C_{\alpha}^{\sigma_{j}}(\rho_{B|\Pi_{i}^{a}})$ has a different definition  depending on the coherence measure used:

\begin{itemize}
\item  \textbf{$l_{1}$-norm:} $C^{\sigma_{j}}_{l_{1}}=\sqrt{\frac{\sum_{i\ne j}\alpha^{2}_{ik_{a}}}{\gamma_{k_{a}}^{2}}}$
\item \textbf{relative entropy:}  $C^{\sigma_{j}}_{re}=\sum_{p=+,-} \lambda_{k_{a}}^{p}\log_{2}\lambda_{k_{a}}^{p}-\beta_{jk_{a}}^{p}\log_{2}\beta_{jk_{a}}^{p}$
\item \textbf{skew information:}  $C^{\sigma_{j}}_{sk}=\frac{(\sum_{i\ne j}\alpha_{ik}^{2})(1-\sqrt{1-(2\lambda_{k_{a}}^{\pm}-1)^{2})}}{\gamma_{ka}^{2}(2\lambda_{ka}^{\pm}-1)^{2}}$
\end{itemize} 
where $\alpha_{ij_{a}}=s_{i}+(-1)^{a}t_{ij}$, $\gamma_{k_{a}}=1+(-1)^{a}r_{k}$, $\lambda_{i_{a}}^{\pm}=\frac{1}{2}\pm\frac{\sqrt{\sum_{j}\alpha^{2}_{ji_{a}}}}{2\gamma_{i_{a}}}$ and $\beta_{ij_{a}}^{\pm}=\frac{1}{2}\pm\frac{\alpha_{ij_{a}}}{2\gamma_{j_{a}}}$.

Following these instructions it is simple to obtain the expressions given by Eqs.(\ref{7},\ref{c2},\ref{c3}) for the NAQCs.

\section{Wave packet description of neutrino oscillations. }
\label{AppendixB}
We briefly review the wave packet approach to neutrino oscillations \cite{Giunti,Giunti1}.
Let us consider a neutrino with flavor $\alpha$, $(\alpha=e,\mu,\tau)$, propagating along $x$ axis:

\begin{equation}
\ket{\nu_{\alpha}}(x,t)=\sum_{j}U_{\alpha j}^{*}\psi_{j}(x,t)\ket{\nu_{j}},
\label{sub1}
\end{equation}
where $U_{\alpha j}$ are the PMNS mixing matrix elements and $\psi_{j}(x,t)$ is the wave function of the mass eigenstate $\ket{\nu_{j}}$ with mass $m_{j}$. By assuming a Gaussian distribution for the momentum of the massive neutrino $\nu_{j}$:

\begin{equation}
\psi_{j}(p)=\bigl(2\pi {\sigma_{p}^{P}}^{2}\bigl)^{-\frac{1}{4}}\exp{-\frac{(p-p_{j})^{2}}{4{\sigma_{p}^{P}}^{2}}}
\label{sub2}
\end{equation}
where $p_{j}$ is the average momentum and $\sigma_{p}^{P}$ is the momentum uncertainty determined by the production process, we can write the wave function as:

\begin{equation}
\psi_{j}(x,t)=\frac{1}{\sqrt{2\pi}}\int dp \hspace{0.1cm}\psi_{j}(p)e^{ipx-iE_{j}(p)t}, 
\label{sub3}
\end{equation}
where  $E_{j}(p)=\sqrt{p^{2}+m_{j}^{2}}$ is the energy.
We suppose that the Gaussian momentum  distribution, Eq. (\ref{sub2}),  is strongly peaked around $p_{j}$, that is, we assume the condition $\sigma_{p}^{P}\ll E_{j}^{2}(p_{j})/m_{j}$. This allows us to approximate the energy with:

\begin{equation}
E_{j}(p)\simeq E_{j} + v_{j}(p-p_{j}),
\label{sub4}
\end{equation}
where $ E_{j}=\sqrt{p_{j}^{2}+m_{j}^{2}}$ is the average energy and $v_{j}=\frac{\partial E_{j}(p)}{\partial p}\biggl|_{p=p_{j}}=\frac{p_{j}}{E_{j}}$ is the group velocity of the wave packet of the massive neutrino $\nu_{j}$.

Using these approximations we can perform an integration on $p$ of  Eq. (\ref{sub3}), obtaining:

\begin{equation}
\psi_{j}(x,t)=\bigl(2\pi {\sigma_{x}^{P}}^{2}\bigl)^{-\frac{1}{4}} \exp \biggl[-iE_{j}t + ip_{j}x - \frac{(x-v_{j}t)^{2}}{4 {\sigma_{x}^{P}}^{2}}\biggl]
\label{sub5}
\end{equation}
where $\sigma_{x}^{P}=\frac{1}{2\sigma_{p}^{P}}$ is the spatial width of the wave packet.

At this point, by substituting  Eq. (\ref{sub5}) in Eq. (\ref{sub1}) it is possible to obtain the density matrix operator by $\rho_{\alpha}(x,t)=\ket{\nu_{\alpha}(x,t)}\bra{\nu_{\alpha}(x,t)}$ which describes the neutrino oscillations in space and time. Although in laboratory experiments it is possible to measure neutrino oscillations in time through the measurement of both the production and detection processes, due to the long time exposure in time of the detectors  it is convenient to consider an average in time of the density matrix operator. In this way $\rho_{\alpha}(x)$ is the relevant density matrix operator and it can be obtained by a gaussian time integration.

In the case of ultra-relativistic neutrinos, it is useful to consider the following approximations: $E_{j}\simeq E + \xi_{P}\frac{m_{j}^{2}}{2E}$, where $E$ is the neutrino energy in the limit of zero mass and $\xi_{P}$ is a dimensionless quantity that depends on the characteristics of the production process, $p_{j}\simeq E-(1-\xi_{P})\frac{m_{j}^{2}}{2E}$ and $v_{j}\simeq 1-\frac{m_{j}^{2}}{2E_{j}^{2}}$.
Considering these approximations,  $\rho_{\alpha}(x)$ becomes:

\begin{equation}
\rho_{\alpha}(x)=\sum_{j,k}U_{\alpha j}^{*}U_{\alpha k} \exp\biggl[-i \frac{\Delta m_{jk}^{2}x}{2E}-\biggl(\frac{\Delta m_{jk}^{2}x}{4 \sqrt{2} E^{2} \sigma_{x}^{P}}\biggl)^{2} - \biggl(\xi_{P}\frac{\Delta m_{jk}^{2}}{4 \sqrt{2} E \sigma_{p}^{P}}\biggl)^{2}\biggl]\ket{\nu_{j}}\bra{\nu_{k}},
\label{sub8}
\end{equation}
where $\Delta m_{jk}^{2}=m_{j}^{2}-m_{k}^{2}$.

Taking into account that the detection process, described by the operator $\mathcal{O}_{\beta}(x-L)$, takes place at a distance $L$ from the origin of the coordinates, the transition probability is given by:

\begin{equation}
P_{\nu_{\alpha}\rightarrow \nu_{\beta}}(L)= \Tr\bigl(\rho_{\alpha}(x)\mathcal{O}_{\beta}(x-L)\bigl)
= \sum_{j,k}U_{\alpha j}^{*}U_{\alpha k}U_{\beta j}^{*}U_{\beta k}\exp\biggl[-2\pi i \frac{L}{L^{osc}_{jk}}-\biggl(\frac{L}{L^{coh}_{jk}}\biggl)^{2}- 2 \pi^{2}(1-\xi)^{2}\biggl(\frac{\sigma_{x}}{L^{osc}_{jk}}\biggl)^{2}\biggl],
\label{sub10}
\end{equation}

where $L^{osc}_{jk}$ is the oscillation length and $L^{coh}_{jk}$  the coherence length, defined by:

\begin{equation}
L^{osc}_{jk}=\frac{4 \pi E}{\Delta m_{jk}^{2}}, \hspace{1cm} L^{coh}_{jk}=\frac{4 \sqrt{2} E^{2}}{|\Delta m_{jk}^{2}|}\sigma_{x},
\label{sub11}
\end{equation}
with $\sigma_{x}^{2}={\sigma_{x}^{P}}^{2} + {\sigma_{x}^{D}}^{2}$ and $\xi^{2}\sigma_{x}^{2}=\xi_{P}^{2}{\sigma_{x}^{P}}^{2} + \xi_{D}^{2}{\sigma_{x}^{D}}^{2}$,where $\sigma^{D}$ is the uncertainty of the detection process and $\xi_{D}$  depends from the characteristics of the detection process. \\

We note that the  wave packet description confirms the standard value of the oscillation length.  The coherence length is the distance beyond which the interference of the massive neutrinos $\nu_{j}$ and $\nu_{k}$ is suppressed.  The last term in the exponential of  Eq. (\ref{sub11}) implies that the interference of the neutrinos is observable only if the localization of the production and detection processes is smaller than the oscillation length.

\end{appendices}

\end{document}